\documentclass[twocolumn,twoside]{IEEEtran}

\usepackage[colorlinks]{hyperref}
\usepackage{multirow}
\usepackage{times}
\usepackage{amsmath,amssymb}
\usepackage{algorithm}
\usepackage{epstopdf}
\usepackage{subfigure}
\usepackage[table,dvipsnames]{xcolor}
\usepackage{algorithm}
\usepackage{algorithmic}
\usepackage{cite}
\usepackage{fancyvrb}
\usepackage[framemethod=TikZ]{mdframed}
\usepackage{multicol,booktabs,tabularx}

\usepackage{bbm}
\newtheorem{theorem}{Theorem}

\newenvironment{proof}{\paragraph*{Proof}}{\hfill$\square$}

\newcommand{\boxedthm}[1]{
\begin{mdframed}[roundcorner=5pt,middlelinewidth=2pt,backgroundcolor=black!10]
\vspace{-1ex}
#1
\end{mdframed}
}

{\begin{list}               
    {$\bullet$ \hfill}{
        \setlength{\leftmargin}{\parindent}
        \setlength{\parsep}{0.04\baselineskip}
        \setlength{\itemsep}{0.5\parsep}
        \setlength{\labelwidth}{\leftmargin}
        \setlength{\labelsep}{0em}}
    }
{\end{list}}

\providecommand{\eref}[1]{\eqref{#1}}  
\providecommand{\cref}[1]{Chapter~\ref{#1}}

\providecommand{\fref}[1]{Figure~\ref{#1}}

\providecommand{\R}{\ensuremath{\mathbb{R}}}

\providecommand{\bydef}{\overset{\text{def}}{=}}

\renewcommand{\vec}[1]{\ensuremath{\boldsymbol{#1}}}

\providecommand{\calL}{\mathcal{L}}
\providecommand{\calM}{\mathcal{M}}

\providecommand{\mA}{\mathbf{A}}
\providecommand{\mB}{\mathbf{B}}

\providecommand{\mD}{\mathbf{D}}

\providecommand{\mH}{\mathbf{H}}
\providecommand{\mI}{\mathbf{I}}
\providecommand{\mJ}{\mathbf{J}}

\providecommand{\vt}{\mathbf{t}}
\providecommand{\vu}{\mathbf{u}}

\providecommand{\vx}{\mathbf{x}}

\providecommand{\vrho}{\vec{\rho}}

\providecommand{\vone}{\vec{1}}

\newcommand{\argmin}[1]{\mathop{\underset{#1}{\mbox{argmin}}}}

\newcommand{\rev}[1]{\textcolor{black}{#1}}
\newcommand{\revv}[1]{\textcolor{black}{#1}}

\title{Scattering and Gathering for Spatially Varying Blurs}
\author{Nicholas~Chimitt,~\IEEEmembership{Member,~IEEE}, Xingguang~Zhang,~\IEEEmembership{Student Member,~IEEE}, \\ Yiheng~Chi,~\IEEEmembership{Student Member,~IEEE}, and Stanley~H.~Chan,~\IEEEmembership{Senior Member,~IEEE}
\thanks{The authors are with the School of Electrical and Computer Engineering, Purdue University, West Lafayette, IN 47907, USA. Corresponding author: Stanley Chan, email: \texttt{stanchan@purdue.edu}.}
\thanks{The work is supported in part by the Intelligence Advanced Research Projects Activity (IARPA) under Contract No. 2022-21102100004, and in part by the National Science Foundation under the grants CCSS-2030570 and IIS-2133032. The views and conclusions contained herein are those of the authors and should not be interpreted as necessarily representing the official policies, either expressed or implied, of IARPA, or the U.S. Government. The U.S. Government is authorized to reproduce and distribute reprints for governmental purposes notwithstanding any copyright annotation therein.}
}

\graphicspath{{pix/}}

\begin{document}
\maketitle

\begin{abstract}
A spatially varying blur kernel $h(\vx,\vu)$ is specified by an input coordinate $\vu \in \R^2$ and an output coordinate $\vx \in \R^2$. For computational efficiency, we sometimes write $h(\vx,\vu)$ as a linear combination of spatially invariant basis functions. The associated pixelwise coefficients, however, can be indexed by either the input coordinate or the output coordinate. While appearing subtle, the two indexing schemes will lead to two different forms of convolutions known as \emph{scattering} and \emph{gathering}, respectively. We discuss the origin of the operations. We discuss conditions under which the two operations are identical. We show that scattering is more suitable for simulating how light propagates and gathering is more suitable for image filtering such as denoising.
\end{abstract}


\section{Introduction}
\rev{In two-dimensional} space, the convolution between an input image $J(\vx)$ and a shift-invariant kernel $h(\vx)$ produces an output image $I(\vx)$ via the well-known integral
\begin{equation}
I(\vx) = \int_{\rev{\R^2}} h(\vx-\vu) J(\vu) \; d\vu.
\label{eq: convolution}
\end{equation}
In this equation, $\vx \in \R^2$ is a two-dimensional coordinate in the output space and $\vu \in \R^2$ is a coordinate in the input space. This definition is ubiquitous in all shift-invariant systems.

If the kernel $h$ is  spatially \emph{varying}, then it is no longer a function of the coordinate difference $\vx-\vu$ but a function of two variables $\vx$ and $\vu$. The resulting kernel $h(\vx,\vu)$ will give the input-output relationship via the integral
\begin{equation}
I(\vx) = \int_{\rev{\R^2}} h(\vx, \vu) J(\vu) \; d\vu,
\label{eq: I varying}
\end{equation}
also known as the superposition integral.

While spatially varying kernels are more difficult to analyze because they cannot be directly handled by Fourier transforms \cite{Chan_2010_a}, they are common in image \emph{formation} and image \emph{processing} which are twins in many situations, such as in the case of kernel estimation \cite{Chakrabarti_2010_a, Zhu_2013_a, Huang_2021_a} or image restoration \cite{Sondhi_1972_a, Suin_2021_a, oConnor_2017_a}. In image formation, spatially varying kernels are used to model how light propagates from the object plane to the image plane. These kernels are known \rev{as point} spread functions (PSFs) which may be spatially varying due to various degradations in the medium or aberrations in the imaging system (such as spherical aberration). In image processing, spatially varying kernels are used to filter the input image for applications such as denoising or interpolation. The spatially varying nature in these situations come from examples such as non-local edge-aware filters where the shapes and orientations of the filters change depending on the image.

The theme of this paper is about the decomposition of the kernel in terms of basis functions. If $h$ is \emph{spatially invariant}, we may express it via the equation
\begin{equation}
h(\vx-\vu) = \sum_{m=1}^M a_{m} \varphi_m(\vx-\vu),
\label{eq: h = sum, invariant}
\end{equation}
where $\{\varphi_1,\varphi_2,\ldots,\varphi_M\}$ are orthogonal basis functions. These functions could be as \rev{simple} as the derivatives of Gaussians, or they can be learned from a dataset of kernels via principal component analysis \rev{(PCA)}. The scalars $\{a_1,a_2,\ldots,a_M\}$ are the basis coefficients, often constructed according the local image statistics or the underpinning physics.

In the case of \emph{spatially varying} kernels, the subject of this paper, the basis representation in \eref{eq: h = sum, invariant} needs to be modified so that it can take the two variables $\vx$ and $\vu$ \rev{into account}. However, there is an ambiguity due to the existence of two options which we call \emph{gathering} and \emph{scattering}:
\begin{align}
\text{(Gathering)}\qquad h(\vx,\vu) &= \sum_{m=1}^M a_{\vx,m} \varphi_m(\vx-\vu), \label{eq: x}\\
\text{(Scattering)}\qquad h(\vx,\vu) &= \sum_{m=1}^M a_{\vu,m} \varphi_m(\vx-\vu).  \label{eq: u}
\end{align}
In both options, the spatially varying kernel $h(\vx,\vu)$ is written as a combination of \emph{invariant} kernels $\{\varphi_1,\varphi_2,\ldots,\varphi_M\}$. These $\varphi_m$'s are spatially invariant, so they can be written as $\varphi_m(\vx-\vu)$. The difference between the two options lies in the coefficient $a_{\vx,m}$ and $a_{\vu,m}$. \rev{Both possibilities have utility; this paper is aimed at describing their appropriateness.}

\textbf{Remark}: Readers may wonder if we can define $h(\vx,\vu)$ using a global $a_{m}$ instead of a pixelwise $a_{\vx,m}$ or $a_{\vu,m}$. If we do so, for example by defining $h(\vx,\vu) = \sum_{m=1}^M a_m \varphi_m(\vx-\vu)$, then $h(\vx,\vu)$ will be invariant because it is a linear combination of invariant basis functions. This will defeat the purpose studying a set of varying kernels. $\square$

At first glance, the two choices above seem subtle to an extent that one may expect a minor difference in terms image quality. However, the two equations have two fundamentally different physical meanings. 
To give readers a preview of the main claims of the paper, we summarize as follows:
\begin{itemize}
\item \textbf{Gathering:} $a_{\vx,m}$ is for image \emph{processing} such as denoising filter, interpolation filter, etc.
\item \textbf{Scattering:} $a_{\vu,m}$ is for image \emph{formation} such as modeling \rev{motion blur or} atmospheric turbulence.
\end{itemize}

The study of time variant systems goes back to classical time-variant filter banks, wavelets, or Kalman filtering/state-space applications \cite{Saleh_1985_a, Sodagar_1994_a, Sodagar_1995_a, Wang_2002_a, Arulampalam_2002_a, Wang_2006_a}. The study of spatially varying kernels as a sum of invariant ones has been performed in works such as \cite{Lauer_2002_a, Flicker_2005_a, Hirsch_2010_a, Arigovindan_2010_a, Mitchel_2020_a, Yanny_2022_a} \rev{as applied models but often do not address the reason for the forward model chosen. The role of this work is to clearly highlight} the difference between the two approximations from the side of modeling and describing where each one is more appropriately applied.

\section{\rev{The Operations} of Gathering and Scattering}

\subsection{Understanding Gathering}
\rev{To present the operation of gathering, we can substitute the basis representation of \eref{eq: x} into \eref{eq: I varying}, resulting in}
\begin{align}
I(\vx)
&\overset{\text{by}\, \eref{eq: I varying}}{=}  \int_{\rev{\R^2}} h(\vx, \vu) J(\vu) \; d\vu \notag \\
&\overset{\text{by}\, \eref{eq: x}}{=}          \int_{\rev{\R^2}} \left(\sum_{m=1}^M a_{\vx,m} \varphi_m(\vx-\vu)\right) J(\vu) \; d\vu \notag\\
&= \sum_{m=1}^M a_{\vx,m}  \left(\int_{\rev{\R^2}} \varphi_m(\vx-\vu) J(\vu) \; d\vu\right). \notag
\end{align}
Recognizing that the integral is a spatially invariant convolution, we can show that
\begin{align}
I(\vx)
&= \underset{\text{linear combination of invariant blurs}}{\underbrace{\sum_{m=1}^M a_{\vx,m} \;\; \underset{\text{invariant blur}}{\underbrace{(\varphi_m \rev{\ast} J)(\vx)}}}}. \label{eq: I x}
\end{align}
Therefore, spatially varying convolution is replaced by a sum of $M$ spatially \emph{invariant} convolutions. Each term in the sum $(\varphi_m \rev{\ast} J)(\vx)$ is the convolution between the image $J(\vu)$ and the basis functions $\varphi_m(\vu)$.

\begin{figure}[h]
	\centering
	\includegraphics[width=\linewidth]{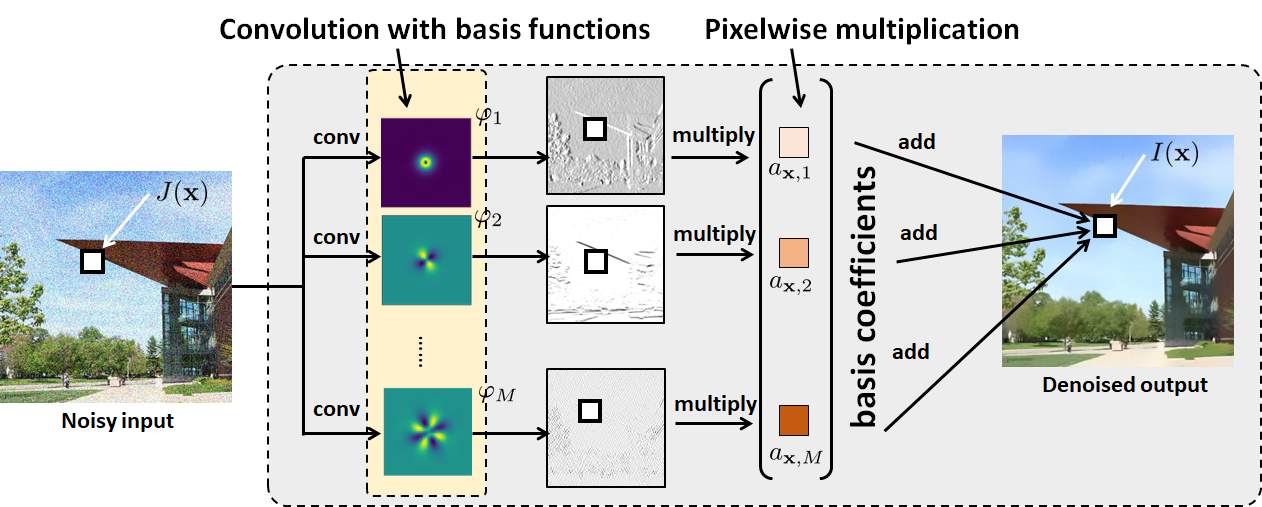}
	\vspace{-3ex}
	\caption{\textbf{Gathering is for Image Processing}. In gathering, we construct a set of convolved images $(\varphi_m \rev{\ast} J)(\vx)$ and \emph{gather} them via pixelwise multiplication. The process offers a significant saving in terms of computation. }
	\label{fig: NLM}
\end{figure}

In terms of computation, we start with the image $J(\vu)$, pre-compute the filtered images $(\varphi_m \rev{\ast} J)(\vx)$, then form a weighted sum of these images to construct the output $I(\vx)$. The weights are pixelwise, and so the resulting blur is spatially varying. \fref{fig: NLM} illustrates how this idea can be implemented. \rev{Because of the ordering of operations, this has also been referred to as convolution-product \cite{Busby_1981_a} in the more mathematical literature}.

The illustration in \fref{fig: NLM} should remind readers of many classical signal processing techniques. For example, the steerable filters by Freeman and Adelson \cite{Freeman_1991_a} employed the idea to detect image edges using a bank of Laplacian filters. The idea was also used in applied mathematics literature. For example, Nagy and O'Leary \cite{Nagy_1998_a} used the idea to decompose the spatially varying kernels using non-overlapped blocks. Nagy and O'Leary's formulation is a special case of our model by making the coefficients $a_{\vx,m}$ as a binary mask. In \cite{Popkin_2010_a}, Popkin et al. used the idea to perform fast computation of spatially varying blurs with the basis functions computed via \rev{PCA}.

\begin{figure}[h]
\centering
\includegraphics[width=0.6\linewidth]{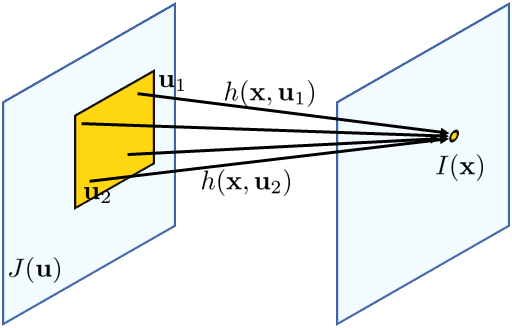}
\caption{The operation derived in \eref{eq: I x} follows the intuitive argument that a neighboring input pixels are mapped to an output. Thus, each $h(\vx,\vu)$ is defined based on the output coordinate $\vx$: At every $\vx$, there is a set of basis functions that define $h(\vx,\vu)$.}
\label{fig: x}
\end{figure}

To readers who prefer the perspective of deep learning, gathering as in \eref{eq: x} is what we call \emph{convolution} in deep neural networks. To see why gathering is identical to the convolution (in deep-learning), we can refer to \fref{fig: x}. Starting with a local neighborhood in the input space, we assign each pixel a weight which is specified by the kernel and map the sum to the output coordinate.

We summarize our findings here:
\boxedthm{
\vspace{1ex}
What is \textbf{gathering \rev{(convolution-product)}}?
\begin{itemize}
\itemsep\setlength{0ex}
\item We apply spatially invarying kernels first, and then combine the results with weights.
\item Consistent with \rev{convolution presented} in Oppenheim and Wilsky \cite{Oppenheim_1996_a}, ``flip, shift, and integrate''.
\item Equivalent to the ``convolution'' in deep neural networks.
\end{itemize}
}

\subsection{Understanding Scattering}
\rev{For scattering}, we follow the same approach as we did for gathering. Substituting \eref{eq: u} into \eref{eq: I varying}, we see that
\begin{align}
I(\vx)
&\overset{\text{by}\, \eref{eq: I varying}}{=} \int_{\rev{\R^2}} h(\vx, \vu) J(\vu) \; d\vu \notag\\
&\overset{\text{by}\, \eref{eq: u}}{=} \int_{\rev{\R^2}} \left(\sum_{m=1}^M a_{\vu,m} \varphi_m(\vx-\vu)\right) J(\vu) \; d\vu \notag\\
&= \sum_{m=1}^M  \left(\int_{\rev{\R^2}} \varphi_m(\vx-\vu) \cdot
\Big(a_{\vu,m} J(\vu)\Big)
\; d\vu\right). \notag
\end{align}
The last equation is trickier to understand. For a fixed index $m$, the product term $a_{\vu,m} J(\vu)$ is a pixelwise multiplication of the coefficient map $a_{\vu,m}$ and the input image $J(\vu)$. Therefore, if we express it using the elementwise multiplication $\odot$, we can write the equation as
\begin{align}
I(\vx) &= \underset{\text{linear combination of invariant blurs}}{\underbrace{\sum_{m=1}^M \bigg( \varphi_m \quad \rev{\ast} \underset{\text{pixelwise weighted input}}{\underbrace{\Big(a_m \odot J \Big)}} \bigg)(\vx)}} \label{eq: I u}.
\end{align}

\rev{Again thinking in terms of computation,} we need to first construct the $M$ coefficient maps $\{a_{\vu,1},\ldots,a_{\vu,M}\}$ where each $a_{\vu,m}$ is a value in an array of the same size as the image $J$. We multiply $J$ with each coefficient map to obtain $M$ ``weighted images'' $\widetilde{J}_m(\vu) = a_{\vu,m}J(\vu)$. Then we convolve these $M$ intermediate images through $(\varphi_m \rev{\ast} \widetilde{J}_m)(\vu)$. Finally, summing over the $M$ images will give us the final result.

\begin{figure}[h]
\centering
\includegraphics[width=\linewidth]{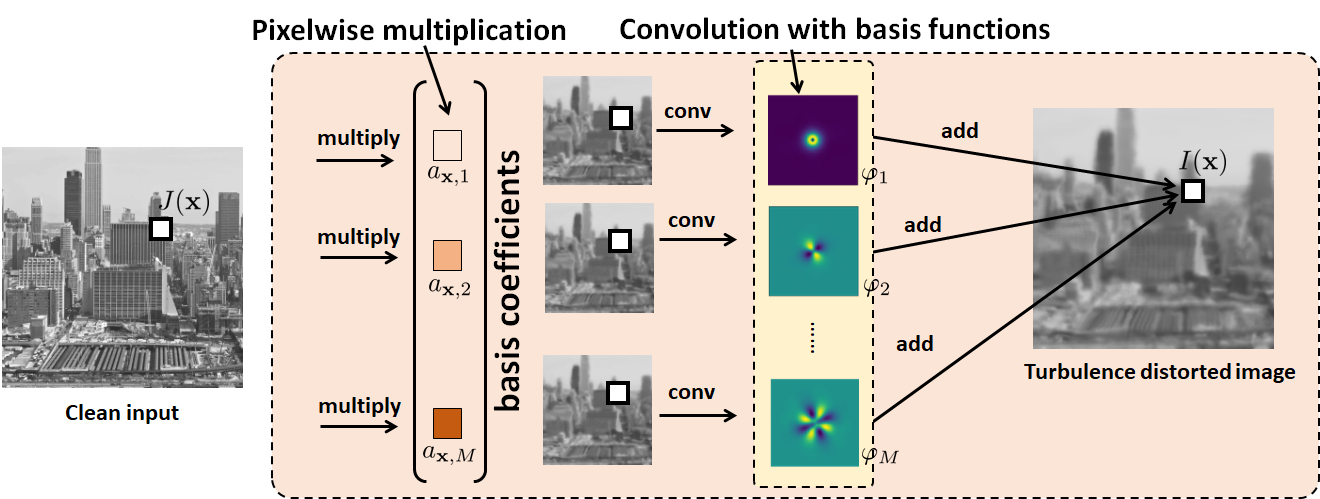}
\vspace{-3ex}
\caption{\textbf{Scattering is for Image Formation}. Here we show a modified version of Mao et al's phase-to-space (P2S) transform \cite{Mao_2021_ICCV} by switching the order of multiplication and convolution. The new formulation follows the derivation in \eref{eq: I u}. We argue that this formulation is the better (and the physically valid) approach when modeling how light propagates from the object plane to the image plane.}
\label{fig: P2S new}
\end{figure}

The process is summarized in \fref{fig: P2S new}. Comparing it with \fref{fig: NLM}, we notice that the difference is \rev{order of multiplication and convolution}. In \fref{fig: NLM}, the pixelwise multiplication is done \emph{after} the convolutions whereas in \fref{fig: P2S new}, it is done \emph{before} the convolutions. Since the two processes involve the same number of convolutions and multiplications, the complexity of executing the computation is identical.

\rev{The role of scattering has been considered in a variety of mathematical studies, such as in Busby et al. \cite{Busby_1981_a} or Escande et al. \cite{Escande_2017_a} in which various bounds can be placed on the approximation of a product-convolution for a general operator \eqref{eq: I varying}. Alger et al. \cite{Alger_2019_a} demonstrated that product-convolution arises naturally in the case of locally-invariant kernels and proposes an adaptive scheme for the coefficients.}

\begin{figure}[h]
\centering
\includegraphics[width=0.6\linewidth]{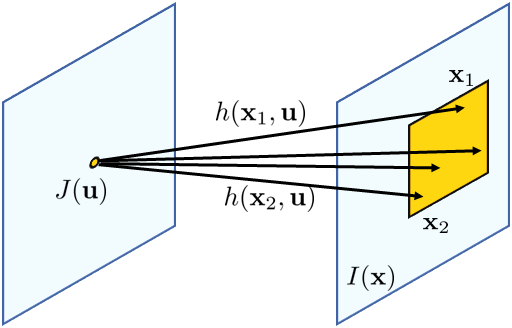}
\caption{The operation behind the Scattering equation: A point source $\delta(\vu)$ emits light which propagates to the object plane. The point spread function (aka the blur kernel $h$) seen in the object plane is centered at $\vu$. At each fixed $\vu$, the pixel values $h(\vx,\vu)$ are determined by how the propagation medium affects the amplitude and phase at coordinate $\vx$.}
\label{fig: u}
\end{figure}

The situation described in \eref{eq: I u} is the mirror of \fref{fig: x}. Instead of having a local neighborhood whose pixels are mapped to a common output coordinate, the input pixel distributes its influence to a local neighborhood in the output space, as shown in \fref{fig: u}. In deep learning, we call it the \emph{transposed convolution}, which is typically used to upsample the features in tasks such as image super-resolution.

We summarize our findings here:
\boxedthm{
\vspace{1ex}
What is \textbf{scattering \rev{(product-convolution)}}?
\begin{itemize}
\itemsep\setlength{0ex}
\item We apply weighted averaging first, and then filter the weighted averages, and finally add.
\item Consistent with how light propagates. See Goodman's \emph{Fourier Optics} \cite{Goodman_2005_FourierOptics} and our next section.
\item Equivalent to the ``transposed convolution'' in deep neural networks.
\end{itemize}
}

\subsection{Conditions for Equivalence}
After elaborating on the computations of gathering and scattering, we now explain the conditions under which the two are equivalent. As expected, the two are \revv{only equivalent in highly special cases.}

Let us look at the equations more carefully through the lens of matrices and vectors. Let $\mH \in \R^{N \times N}$ be the matrix representation of the spatially varying blur kernel $h(\vx,\vu)$. We assume that there is a set of \emph{circulant} matrices $\mH_1,\mH_2,\ldots,\mH_M$ representing the set of $M$ spatially invariant basis functions $\{\varphi_1,\varphi_2,\ldots,\varphi_M\}$.

We consider two sets of diagonal matrices. For every index $m$, we define
\begin{align*}
\mD^{\rev{(g)}}_m =
\text{diag}\left\{
\begin{bmatrix}
a_{\vx_1,m} \\
\vdots \\
a_{\vx_N,m}
\end{bmatrix}\right\}, \;\;
\mD^{\rev{(s)}}_m =
\text{diag}\left\{
\begin{bmatrix}
a_{\vu_1,m} \\
\vdots \\
a_{\vu_N,m}
\end{bmatrix}\right\}
\end{align*}
Then, the gathering and scattering equations are
\begin{align}
(\text{Gathering}):  \qquad \mH^{\rev{(g)}} &= \sum_{m=1}^M \mD_m^{\rev{(g)}} \mH_m \label{eq: gathering matrix}\\
(\text{Scattering}): \qquad \mH^{\rev{(s)}} &= \sum_{m=1}^M \mH_m\mD_m^{\rev{(s)}}  \label{eq: scattering matrix}
\end{align}
In other words, the difference lies in how we order the diagonal matrices and the spatially invariant convolution matrices. \revv{We now present a theorem that can be used to show gathering and scattering are mutually exclusive in a wide range of cases.}
\revv{\begin{theorem}
\label{thm: commutative}
Let $\mH \in \R^{N \times N}$ be a circulant matrix formed by $[h_0, \ldots, h_{N-1}]$ with at least two consecutive non-zero elements,  $h_m \neq 0$ and $h_{m+1} \neq 0$. Let $\mA = \text{diag}[a_0,\ldots,a_{N-1}]$ and $\mB = \text{diag}[b_0,\ldots,b_{N-1}]$ be two diagonal matrices. Then, $\mA\mH = \mH\mB$ if and only if $\mA = \mB = \lambda \mI$ for some constant $\lambda$ where $\mI$ is the identity matrix.
\end{theorem}
}

\begin{proof}
\revv{
For the forward direction, we begin by writing out the $(i,j)$th element of each matrix product
\begin{align*}
    [\mA \mH]_{ij} &= a_i h_{ij} = a_i h_{(j-i)\!\mod N} , \\
    [\mH \mB]_{ij} &= b_j h_{ij} = b_j h_{(j-i)\!\mod N} .
\end{align*}
Suppose $h_m \neq 0$, then for any pair $(i,j)$ in the set $\calM = \{(i^*, j^*)\vert (j^* - i^*)\!\mod N = m\}$, the following holds true: $a_i h_m = b_j h_m \implies a_i = b_j.$
The set $\calM$ can be written in terms of $m$ as $\calM = \{(0, m), (1, (m+1)\!\mod N), \ldots, (N-1, (m+N-1)\!\mod N) \}$, with a similar set $\calM'$ for $m+1$ corresponding to $h_{m+1} \neq 0$. Through $\calM$ and $\calM'$ we have $a_k = b_{(m+k)\mod\!N}$ and $a_l = b_{(m+1+l)\mod\!N}$.}

\revv{Observing that $a_0 = b_{m\!\mod N} = b_{(m+1)\!\mod N}$ and in general $a_k = b_{(m+k)\!\mod N} = b_{(m+k+1)\!\mod N}$, we set a sequence $p_k = (m + k) \!\mod N$ for $k = 0, \ldots N-1$. Since it can be observed $p_k$ does not repeat within a cycle of $N$, we can establish $b_{p_0} = \cdots = b_{p_{N-1}}$ $\implies$ $b_0 = \cdots = b_{N-1}$ $\implies$ $a_0 = \cdots = a_{N-1}$, and finally that all $a_i$ and $b_j$ are equal.}

\revv{For the reverse direction, $\mA = \mB = \lambda \mI$ can be used to show $\mA \mH = \lambda \mI \mH = \lambda \mH = \mH \lambda \mI = \mH \mB$.}
\end{proof}

The result of the previous theorem implies that if we have a convolution matrix $\mH_m$ (which is circulant by definition) \revv{that satisfies the consecutive non-zero property,} for the scattering and gathering operations to be equivalent, we need
\begin{equation*}
\mD^{\rev{(g)}}_m\mH_m = \mH_m \mD^{\rev{(s)}}_m, \quad \text{for all } m.
\end{equation*}
Theorem~\ref{thm: commutative} asserts that we need $\mD^{\rev{(g)}}_m = \mD^{\rev{(s)}}_m = \lambda \mI$. But if $\mD^{\rev{(g)}}_m = \mD^{\rev{(s)}}_m = \lambda \mI$, then the underlying blur must be spatially invariant.

Another \revv{consequence} of Theorem~\ref{thm: commutative} is that
\begin{equation}
\sum_{m=1}^M \mD_m^{\rev{(g)}} \mH_m \not= \sum_{m=1}^M \mH_m\mD_m^{\rev{(s)}}.
\end{equation}
Therefore, the gathering and scattering equations \eref{eq: x} and \eref{eq: u} are \emph{mutually exclusive} \revv{under the conditions stated in Theorem~\ref{thm: commutative}}. If we say that $h(\vx,\vu)$ can be exactly represented by the gathering equation, then there will be an approximation error when representing $h(\vx,\vu)$ using the scattering equation, and vice versa.

\boxedthm{
\vspace{1ex}
Under what \textbf{conditions} would scattering = gathering?
\begin{itemize}
\itemsep\setlength{0ex}
\item When the underlying blur is spatially invariant \revv{or does not satisfy the consecutive non-zero property.}
\item Scattering and gathering are mutually exclusive. We cannot simultaneously have \eref{eq: x} and \eref{eq: u} for a spatially varying blur. If one is the correct representation, the other will have approximation error.
\end{itemize}
}

\subsection{Normalization}
When performing a convolution, it is \rev{often} necessary to ensure that the image intensity is not amplified or attenuated due to an improper normalization. For example, in a spatially invariant blur, we almost always require that
\begin{equation*}
\int_{\rev{\R^2}} h(\vu) d\vu = 1,
\end{equation*}
assuming that $h(\vu) \ge 0$ for all $\vu$. Otherwise, if the integral is less than unity, the resulting (convolved) image will appear to be dimmer. Translated to matrices and vectors, this is equivalent to $\mH\vone = \vone$ for an all-one vector $\vone$, assuming that $\mH$ is circulant.

Suppose that we have a sequence of spatially invariant blurs $\mH_1,\mH_2,\ldots,\mH_M$ satisfying the property that $\mH_m\vone = \vone$ for all $m$. We want the diagonal matrices $\mD_1^{\vx},\mD_2^{\vx},\ldots,\mD_m^{\rev{(g)}}$ to be defined in such a way that the gathering equation \eref{eq: gathering matrix} will give us
\begin{align*}
\vone \quad \overset{\text{(we want)}}{=} \quad \mH^{\vx}\vone
&= \left(\sum_{m=1}^M \mD_m^{\rev{(g)}} \mH_m\right)\vone = \sum_{m=1}^M \mD_m^{\rev{(g)}} \vone.
\end{align*}
Therefore, as long as we can ensure that the sum of the $M$ diagonal matrices $\{\mD_m^{\rev{(g)}} \,|\, m = 1,\ldots,M\}$ is a vector of all one's, we are guaranteed to have $\mH^{\vx}$ to have unit rows. Converting this into the basis representation, it is equivalent to asking
\begin{equation}
\sum_{m=1}^M a_{\vx,m} = \vone, \qquad \text{for all } m,
\end{equation}
which is reasonably easy to satisfy. For implementation, if $\mH^{\vx}$ does not have rows sum to the unity such that $\mH^{\vx}\vone \not= \vone$, the simplest approach is to define a diagonal matrix $\mD$ such that $\mD^{-1}\mH^{\vx}\vone = \vone$. From the derivations above, it is clear that the diagonal matrix should have the elements
\begin{equation}
\mD = \text{diag}\left\{\sum_{m=1}^M \mD_m^{\rev{(g)}} \mH_m \vone\right\}.
\end{equation}
Therefore, the overall operation applied to an image is
\begin{equation}
\widehat{\mI} = \text{diag}\left\{\sum_{m=1}^M \mD_m^{\rev{(g)}} \mH_m \vone\right\}^{-1} \left(\sum_{m=1}^M \mD_m^{\rev{(g)}} \mH_m \mJ\right),
\label{eq: normalize gathering}
\end{equation}
where $\mJ \in \R^N$ is the vector representation of the input, and $\widehat{\mI} \in \R^N$ is the output.

The normalization of the scattering equation is more complicated because the diagonal matrices do not commute with the spatially invariant blurs, and so we are not able to simplify the equation:
\begin{align*}
\vone \quad \overset{\text{(we want)}}{=} \quad \mH^{\vu}\vone
&= \sum_{m=1}^M \mH_m (\mD_m^{\rev{(s)}} \vone).
\end{align*}
The only work around solution is to define a diagonal matrix $\mD$ such that $\mD^{-1}\mH^{\vu}\vone = \vone$. This would require that
\begin{equation}
\mD = \text{diag}\left\{\sum_{m=1}^M \mH_m \mD_m^{\rev{(s)}} \vone\right\}.
\end{equation}
In other words, when computing the scattering equation and if normalization is required, then we need to compute the set of convolutions \emph{twice}: once for the input, and once for the normalization constant,
\begin{equation}
\widehat{\mI} = \text{diag}\left\{\sum_{m=1}^M \mH_m \mD_m^{\rev{(g)}} \vone\right\}^{-1} \left(\sum_{m=1}^M \mH_m  \mD_m^{\rev{(g)}} \mJ\right).
\label{eq: normalize scattering}
\end{equation}
To summarize the normalization:
\boxedthm{
\vspace{1ex}
If \textbf{normalization} is needed, then
\begin{itemize}
\itemsep\setlength{0ex}
\item For gathering: Either ensure $\sum_{m=1}^M a_{\vx,m} = 1$ for all $\vx$, or perform the calculation in \eref{eq: normalize gathering}.
\item For scattering: Perform the calculation in \eref{eq: normalize scattering}. The cost is twice the number of convolutions than gathering.
\end{itemize}
}

\section{Origin of the Scattering Equation}
\label{sec: origin}
\begin{figure}[h]
\centering
\includegraphics[width=0.85\linewidth]{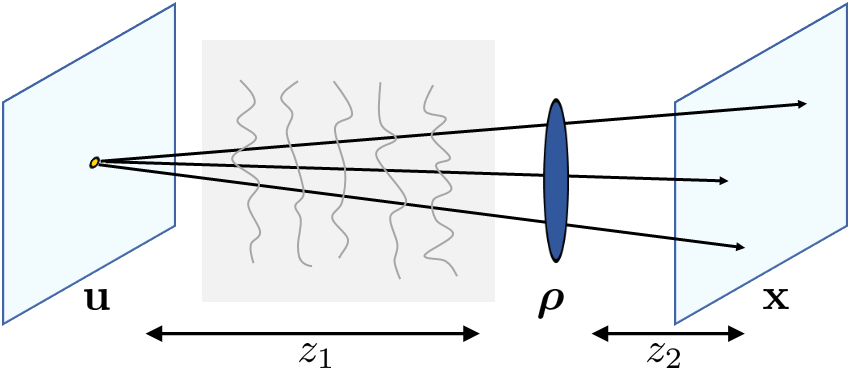}
\caption{The coordinate system of a typical optical system, in the presence of a random medium.}
\label{fig: physics}
\end{figure}

In this section, we explain the origin of the scattering equation from a physics point of view. To make our discussions concrete, the running example we use is a point source propagating a random medium as shown in \fref{fig: physics}.

We follow the coordinate system defined \rev{in} \fref{fig: physics}. The object plane uses the coordinate $\vu \in \R^2$. We can think of it as the \emph{input} coordinate. As light propagates through the random medium, it reaches at the aperture of a lens. The coordinate on the lens is denoted by $\vrho \in \R^2$. \rev{The image plane uses coordinate} $\vx \in \R^2$, which is also the \emph{output} coordinate. Deriving the PSF equation from the Rayleigh-Sommerfeld integral will be too lengthy for our paper. Therefore, we skip the derivation and refer the readers to \cite[Ch. 4 \& 5]{Goodman_2005_FourierOptics} \rev{or \cite{Chan_2023_book}.} Our approach is to highlight the four components of \fref{fig: physics}.

\textbf{Source to Aperture.} The propagation of a point source from the object plane to the aperture, in the absence of the random medium, is characterized by the free-space propagation. The electromagnetic field defined upon the aperture is given by \cite[Eq 5-25]{Goodman_2005_FourierOptics}
\begin{equation}
U(\vu,\vrho) = \frac{1}{j \lambda z_1} \exp\left\{j\frac{k}{2z_1}|\vu - \vrho|^2\right\},
\label{eq: source to lens}
\end{equation}
where $k = 2\pi/\lambda$ is the wave number, and $z_1$ is the distance from the source to the aperture. The notation $|\vu - \vrho|$ denotes the Euclidean distance between the two coordinates $\vu$ and $\vrho$. This equation describes a parabolic wave propagating outward from $\vu$. The farther apart $\vu$ and $\vrho$ is, the weaker the field $U(\vu,\vrho)$ will become.

\textbf{Aperture and lens.} Right at the lens, the incident field will be imparted by the pupil function of the lens and its phase response. For a lens with a focal length of $f$, the field at the exit of the aperture is \cite[Eq 5-26]{Goodman_2005_FourierOptics}
\begin{equation}
U'(\vu,\vrho) = U(\vu,\vrho) P(\vrho)\exp\left\{j\frac{k}{2f}|\vrho|^2\right\},
\end{equation}
where $P(\vrho)$ is the pupil function, typically chosen to be a circular indicator function.

\textbf{Aperture to image.} When the incident field exits the lens, it propagates via Fresnel diffraction to the image plane. Referring to \cite[Eq 5-27]{Goodman_2005_FourierOptics}, we can show that
\begin{equation}
\underset{=h(\vx,\vu)}{\underbrace{U''(\vx,\vu)}} = \frac{1}{j\lambda z_2} \int_{\rev{\R^2}} U'(\vu,\vrho)\exp\left\{ j \frac{k}{2z_2} |\vx - \vrho|^2 \right\} \, d\vrho.
\end{equation}
Notice that the final electromagnetic field $U''(\vx,\vu)$ arriving at the image plane is originated from a point source. As such, $U''(\vx,\vu)$ is the point spread function $h(\vx,\vu)$.

The PSF $h(\vx,\vu)$ can be expressed (with some approximation) as \cite[Eq 5-36]{Goodman_2005_FourierOptics}:
\begin{equation}
h(\vx,\vu) =
\underset{=h(\vx-\vu) \,\, \text{if $S = 1$}}{\underbrace{\kappa \int_{\rev{\R^2}} P(\vrho) \exp\left\{ -j \frac{k}{z_2} (\vx-S\vu)^T \vrho \right\} d\vrho}},
\end{equation}
where $S = -z_2/z_1$ is the magnification factor \rev{and $\kappa = 1/(\lambda^2 z_1z_2)$}. If, for simplicity, we assume $z_1 = -z_2$ so that $S = 1$, then $h(\vx,\vu)$ is completely characterized by the coordinate difference $\vx-\vu$. This will give us $h(\vx,\vu) = h(\vx-\vu)$, and so $h(\vx,\vu)$ represents a spatially \emph{invariant} kernel.

\textbf{Random medium.} The fourth element we need to discuss, which is also the source of the problem, is the random medium. The random medium introduces a random amplitude and phase distortion as
\begin{equation}
R_{\vu}(\vrho) =
\underset{\text{amplitude}}{\underbrace{A_{\vu}(\vrho)}} \; \times \;
\exp\{-j\underset{\text{\rev{phase}}}{\underbrace{\phi_{\vu}(\vrho)}}\}.
\label{eq: random medium}
\end{equation}
Notice that in this definition, the distortion has a coordinate pair $(\vu,\vrho)$. \rev{The position $\vx$ is not present and has no impact at this stage.}
\rev{Since} the impact of the random medium takes place from the source-to-aperture, \rev{we append} \eref{eq: random medium} to \eref{eq: source to lens} \rev{to find}
\begin{align*}
U(\vu,\vrho)
&= \frac{A_{\vu}(\vrho)}{j \lambda z_1} \exp\left\{-j\phi_{\vu}(\vrho)\right\} \exp\left\{j\frac{k}{2z_1}|\vu - \vrho|^2\right\}.
\end{align*}
Consequently, the point spread function is
\begin{align}
h(\vx,\vu)
&= \rev{\kappa} \int_{\rev{\R^2}}
\rev{g_{\vu}(\vrho)} \exp\left\{ -j \frac{k}{z_2} (\vx-\vu)^T \vrho \right\} d\vrho
\label{eq: h, general 01}
\end{align}
where again we assumed $S = 1$ \rev{and defined}
\begin{align*}
g_{\vu}(\vrho) \bydef
A_{\vu}(\vrho) \exp\left\{-j\phi_{\vu}(\vrho)\right\} P(\vrho).
\end{align*}

Therefore, \eref{eq: h, general 01} can be seen as the Fourier transform of the random medium and the pupil function via
\begin{align}
h(\vx,\vu)
&= \text{Fourier}\Big\{ g_{\vu}(\vrho) \Big\} \Big|_{\frac{\vx-\vu}{\lambda z_2}},
\end{align}
where we specify that the transform is evaluated at the coordinate $(\vx-\vu)/(\lambda z_2)$. \rev{This indicates that the} PSF $h(\vx,\vu)$ \rev{is, in part, written as} a function of $\vx-\vu$. However, since $g_{\vu}(\vrho)$ is indexed by $\vu$, the resulting PSF $h(\vx,\vu)$ should \rev{inherit} the index $\vu$. This will give us
\begin{align}
h(\vx,\vu)
&= \text{some function of $(\vx-\vu)$, index by $\vu$}, \notag\\
&= \sum_{m=1}^M a_{\vu,m} \varphi_{m}(\vx-\vu),
\end{align}
where in the last step we use the linear combination of basis functions as the model for such $h(\vx,\vu)$. The basis function $\varphi_m$ here captures the spatial invariance, whereas the coefficients $a_{\vu,m}$ capture the spatially varying indices.

\rev{This derivation explains why} in optics simulation, such as imaging through random \rev{media}, \rev{one} must follow the \emph{scattering} equation if we choose to represent the PSF using a set of spatially invariant basis functions. This is due to the \emph{source} location determining the response of the system. \rev{This extends to other optical distortions where the source location parameterizes the system error.}

\section{Do These Actually Matter?}
The question to ask now is: given the gathering equation and the scattering equation, does it really matter if we choose the ``wrong'' one? The goal of this section is to answer this question through a few examples.

\subsection{Scattering Works for Optical Simulation}
\label{sec: ex1}
\rev{Our first example considers} the problem of simulating the resulting image given an \rev{incoherent} light source. The light source $J(\vu)$ we consider here consists of two delta functions:
\begin{equation*}
J(\vu) = \delta(\vu+\Delta) + \delta(\vu - \Delta),
\end{equation*}
where $\Delta$ is a small displacement. For convenience, we define a plane with two halves; \rev{relative to the separation,} $\delta(\vu+\Delta)$ is on the left and $\delta(\vu-\Delta)$ is on the right.

\begin{figure}[h]
\centering
\begin{tabular}{cc}
\hspace{-1ex}\includegraphics[width=0.48\linewidth]{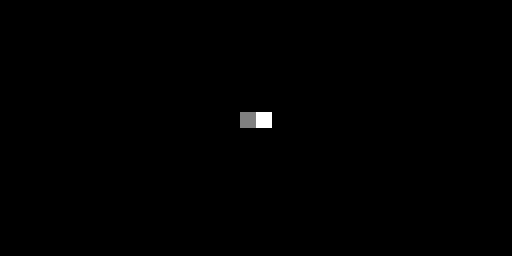}&
\hspace{-2ex}\includegraphics[width=0.48\linewidth]{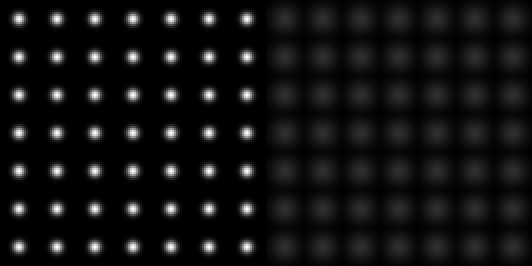}\\
\small{(a) Input $J(\vu)$} & \small{(b) Blurs $h(\vx,\vu)$}\\
    & \scriptsize{(illustrated over a grid of points)}
\end{tabular}
\vspace{-2ex}
\caption{Visualization of an example with (a) $J(\vu)$ and (b) a grid of spatially varying blur kernels.}
\label{fig: Fig_06ab}
\end{figure}

Imagine that in front of the light source, we put two transparent sheets with different phase profiles (which can be engineered using a meta-material\rev{)}. This will give us a spatially varying blur kernel $h(\vx,\vu)$. \rev{By design of the transparent sheet,} if light is emitted on the left hand side the blur \rev{has} a smaller radius; if the light is emitted on the right hand side, then the blur \rev{has} a larger radius. Thus, we write
\begin{equation}
h(\vx,\vu) =
\begin{cases}
\rev{\frac{1}{2\pi \sigma_1^2} e^{-\frac{\|\vx-\vu\|^2}{2\sigma_1^2}}} \bydef \varphi_1(\vx-\vu), &\, \rev{\vu \in \text{left}},\\
\rev{\frac{1}{2\pi \sigma_2^2} e^{-\frac{\|\vx-\vu\|^2}{2\sigma_2^2}}} \bydef \varphi_2(\vx-\vu), &\, \rev{\vu  \in \text{right}},
\end{cases}
\label{eq: Example h}
\end{equation}
where $\sigma_1 < \sigma_2$. \fref{fig: Fig_06ab}(b) illustrates these spatially varying blur kernels. For visualization purposes, we show only the PSFs at a grid of points. In reality, the PSFs \rev{are} defined continuously over $\vu$.

\begin{figure}[h]
\centering
\includegraphics[width=0.8\linewidth]{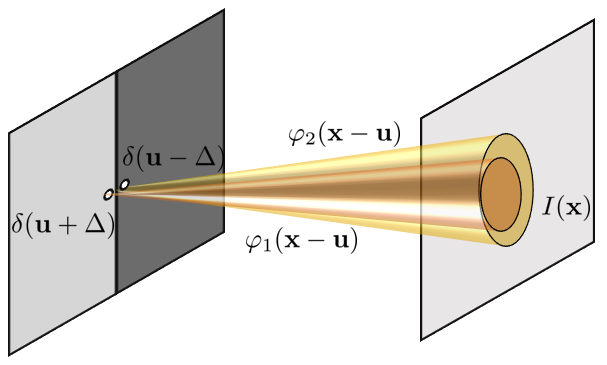}
\caption{Thought experiment with two points on the object plane, diffracting through two different metasurfaces. The resulting image should, in principle, be one superimposed diffraction pattern.}
\label{fig: Figure08_example}
\vspace{-2ex}
\end{figure}

Before we do any calculus, we can perform a thought experiment. \fref{fig: Figure08_example} illustrates a hypothetical experimental setup. On the object plane there are two points emitting light through a meta surface with two different phase profiles. As the light propagates outward from the source through diffraction, the electromagnetic fields superimpose over each other. When the light reaches the aperture, the two diffraction patterns overlap. Therefore, the resulting image $I(\vx)$, without any calculation, should be one big diffraction pattern. It is impossible to obtain a sharp cutoff and two diffraction patterns.

\begin{figure}[h]
\centering
\begin{tabular}{cc}
\hspace{-1ex}\includegraphics[width=0.48\linewidth]{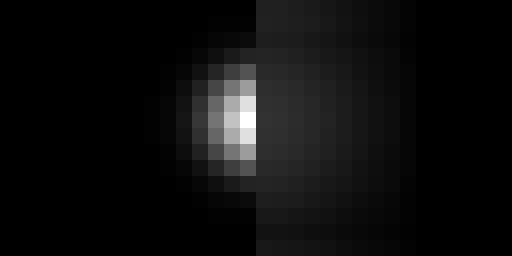}&
\hspace{-2ex}\includegraphics[width=0.48\linewidth]{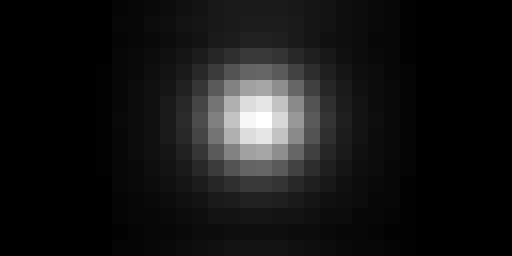}\\
\small{(a) $I^{\text{gather}}(\vx)$} & \small{(b) $I^{\text{scatter}}(\vx)$}
\end{tabular}
\caption{Comparison between gathering and scattering for the setup in \fref{fig: Figure08_example}. Notice that for this optical experiment, we should expect the resulting image to contain one big diffraction pattern. However, only the scattering equation demonstrates this.}
\label{fig: Fig_06cd}
\end{figure}

\rev{With this in mind, we can write the PSF in terms of scattering and gathering as}
\begin{align}
h^{\text{gather}}(\vx,\vu)
&=
\underset{=a_{\vx,1}}{\underbrace{\mathbb{I}\{\vx \in \text{left}\}}} \times \varphi_1(\vx-\vu) \notag \\
&\qquad
+\underset{=a_{\vx,2}}{\underbrace{\mathbb{I}\{\vx \in \text{right}\}}} \times \varphi_2(\vx-\vu),
\label{eq: Example Gather} \\
h^{\text{scatter}}(\vx,\vu)
&=
\underset{=a_{\vu,1}}{\underbrace{\mathbb{I}\{\vu \in \text{left}\}}} \times \varphi_1(\vx-\vu) \notag \\
&\qquad
+\underset{=a_{\vu,2}}{\underbrace{\mathbb{I}\{\vu \in \text{right}\}}} \times \varphi_2(\vx-\vu),
\label{eq: Example Scatter}
\end{align}
where we replaced $a_{\vx,m}$ by $a_{\vu,m}$ and $\mathbb{I}\{\cdot\}$ is the indicator function. By comparing the gathering equation \eref{eq: Example Gather} and the scattering equation \eref{eq: Example Scatter} with the original spatially varying $h(\vx,\vu)$ in \eref{eq: Example h}, it is clear that only the scattering equation will match with the original $h(\vx,\vu)$ because they are both indexed by $\vu$. However, to confirm that this is indeed the case, it would be helpful to look at the resulting image, as illustrated in \fref{fig: Fig_06cd}. 

\rev{We can also mathematically describe the images through use of the kernel $h(\vx,\vu)$ in either case in the superposition integral. The gathering kernel results in
\begin{align}
I^{\text{gather}}(\vx)
&= \int \left(a_{\vx,1}\varphi_1(\vx-\vu) + a_{\vx,2}\varphi_2(\vx-\vu)\right) J(\vu) \,d\vu. \notag \\
&=
\begin{cases}
\varphi_1(\vx+\Delta) + \varphi_1(\vx-\Delta),   &\quad \rev{\vx \in \text{left}},\\
\varphi_2(\vx+\Delta) + \varphi_2(\vx-\Delta),   &\quad \rev{\vx \in \text{right}}.
\end{cases}
\label{eq: Example 1 Gather}
\end{align}
If we draw $I^{\text{gather}}(\vx)$, we will obtain the figure shown in \fref{fig: Fig_06cd}(a). For scattering, we can carry out the same derivation and show that
\begin{align}
I^{\text{scatter}}(\vx)
&= \int_{-\infty}^{\infty} \left(a_{\vu,1}\varphi_1(\vx-\vu) + a_{\vu,2}\varphi_2(\vx-\vu)\right) J(\vu) \,d\vu \notag \\
&= \varphi_1(\vx+\Delta) + \varphi_2(\vx-\Delta).
\label{eq: Example 1 Scatter}
\end{align}
Similar to gathering, }if we draw the resulting image, we will obtain the figure shown in \fref{fig: Fig_06cd}(b). This is consistent with what we expect \rev{from} \fref{fig: Figure08_example} and the theoretical derivation in Section \ref{sec: origin}.

\subsection{Gathering Works for Image Filtering}
\label{sec: ex2}
\begin{figure}[h]
\centering
\begin{tabular}{cc}
\hspace{-1ex}\includegraphics[width=0.48\linewidth]{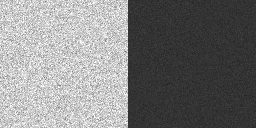}&
\hspace{-2ex}\includegraphics[width=0.48\linewidth]{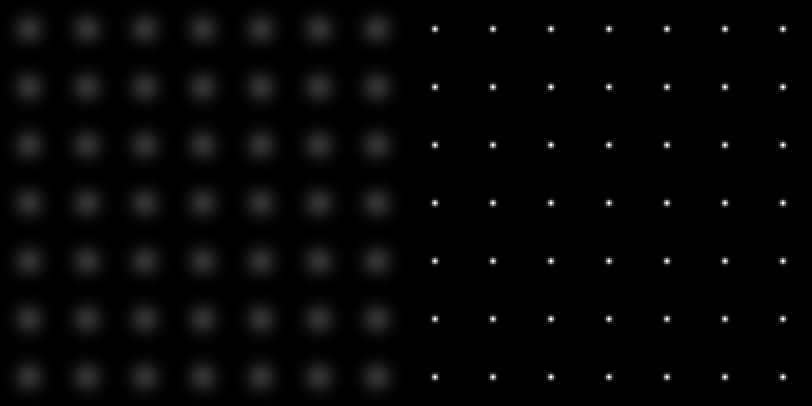}\\
\small{(a) Input $J(\vu)$} & \small{(b) Denoising filter $h(\vx,\vu)$}\\
    & \scriptsize{(illustrated over a grid of points)}
\end{tabular}
\vspace{-2ex}
\caption{Thought experiment of two noisy regions in an image. To denoise this image, ideally we would want to apply to different filters with a sharp boundary at the transition. }
\label{fig: Fig_8 input noisy}
\end{figure}

In the second example, we consider the problem of \emph{image filtering}. The scenario is that we are given a noisy image $J(\vu)$ that contains two regions:
\begin{align}
J(\vu)
&=
\begin{cases}
\theta_1 + \text{Gauss}(0,\sigma_1^2), &\qquad \rev{\vu \in \text{left}},\\
\theta_2 + \text{Gauss}(0,\sigma_2^2), &\qquad \rev{\vu \in \text{right}},
\end{cases}
\end{align}
with two signal levels $\theta_1$ and $\theta_2$, and two noise standard deviations $\sigma_1$ and $\sigma_2$ such that $\sigma_1 > \sigma_2$. In this equation, $W_1(\vu) \sim \text{Gauss}(0,\sigma_1^2)$ and $W_2(\vu) \sim \text{Gauss}(0,\sigma_2^2)$ \rev{denote Gaussian noise}. For illustration, we show in \fref{fig: Fig_8 input noisy}(a) the case where $\theta_1 = 0.8$, $\theta_2 = 0.2$, $\sigma_1 = 0.1$ and $\sigma_2 = 0.02$.

\begin{figure}[h]
\centering
\includegraphics[width=0.8\linewidth]{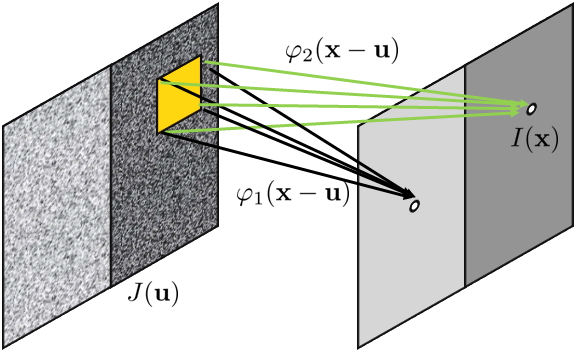}
\caption{Thought experiment with two noisy half-planes. As we perform the denoising step, we would hope that the sharp boundary is preserved.}
\label{fig: Figure09_example}
\vspace{-2ex}
\end{figure}

To denoise this image, we consider the simplest approach assuming that we \emph{knew} the partition of the two regions. Suppose that we want to denoise the left side. Since we know that the noise is stronger, we shall apply a stronger filter. As illustrated in \fref{fig: Figure09_example}, for this filter to be effective along the boundary, we should apply a mask \emph{after} the filtering is done.

The spatially varying filter we propose here takes the form
\begin{align*}
h(\vx,\vu)
&=
\begin{cases}
\frac{1}{\sqrt{2\pi s_1^2}} \rev{e^{-\frac{\|\vx-\vu\|^2}{2s_1^2}}} \bydef \varphi_1(\vx-\vu), & \rev{\vx \in \text{left}},\\
\frac{1}{\sqrt{2\pi s_2^2}} \rev{e^{-\frac{\|\vx-\vu\|^2}{2s_2^2}}} \bydef \varphi_2(\vx-\vu), & \rev{\vx \in \text{right}},
\end{cases}
\end{align*}
where we assume that $s_1 > s_2$. We are careful about the index in this equation, remarking that the conditions are applied to $\vx$ instead of $\vu$. We will illustrate what will happen if the conditions are applied to $\vu$.

The gathering and the scattering equation for this example are identical to those in \eref{eq: Example Gather} and \eref{eq: Example Scatter}. Most importantly, the coefficients $a_{\vx,m}$ and $a_{\vu,m}$ are binary masks indicating whether the pixel $\vx$ (or $\vu$) is on the left / right hand side.

\begin{figure}[h]
\centering
\begin{tabular}{cc}
\hspace{-1ex}\includegraphics[width=0.48\linewidth]{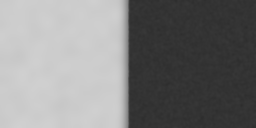}&
\hspace{-2ex}\includegraphics[width=0.48\linewidth]{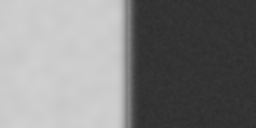}\\
(a) $I^{\text{gather}}(\vx)$ & (b) $I^{\text{scatter}}(\vx)$
\end{tabular}
\vspace{-2ex}
\caption{Comparison between gathering and scattering for the setup in \fref{fig: Fig_8 input noisy}. Notice that for this denoising experiment, the better method should produce a sharp transition along the boundary.}
\label{fig: Fig_9 result}
\vspace{-4ex}
\end{figure}

The resulting images $I^{\text{gather}}(\vx)$ and $I^{\text{scatter}}(\vx)$ follow a similar derivation as in \eref{eq: Example 1 Gather} and \eref{eq: Example 1 Scatter}. \rev{It can be shown that
the two possible approximations reduce to
\begin{align}
    I^{\text{gather}}(\vx) &=
\begin{cases}
(J \ast \varphi_1)(\vx), &\quad \vx < 0,\\
(J \ast \varphi_2)(\vx), &\quad \vx \ge 0,
\end{cases} \\
I^{\text{scatter}}(\vx) &= [(J \times \mathbb{I}\{\vx \in \text{left}\} ) \rev{\ast} \varphi_1](\vx) \notag \\ & \quad + [(J \times \mathbb{I}\{\vx \in \text{right}\}) \rev{\ast} \varphi_2](\vx).\label{eq: Example 2 scatter}
\end{align}
}\rev{This shows that} gathering contains a sharp transition in its output while scattering is a sum of the two convolutions over partitions. We visualize the results in \fref{fig: Fig_9 result}. While both of them can offer denoising to some extent, the gathering approach handles the boundary much better because the masking is performed \emph{after} the filtering. If the masking is performed before the filtering, then \eref{eq: Example 2 scatter} tells us that we are summing two convolutions of the same image. Therefore, the edge is blurred.

\begin{figure}[h]
\centering
\begin{tabular}{cc}
\hspace{-1ex}\includegraphics[width=0.4\linewidth]{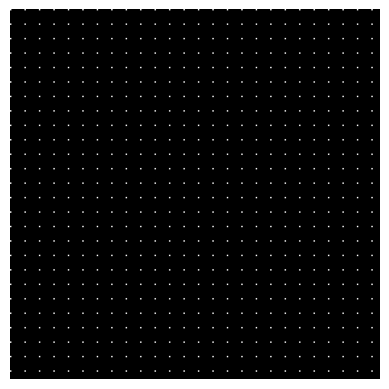}&
\hspace{-2ex}\includegraphics[width=0.4\linewidth]{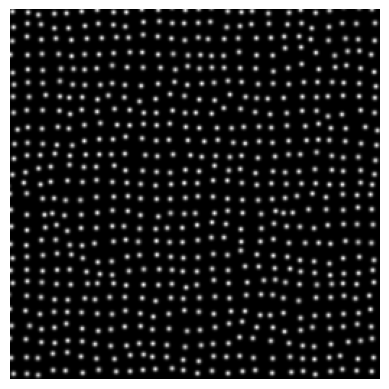}\\
\small{(a) Input} & \small{(b) Output}
\end{tabular}
\caption{\rev{Visualizing the effect of our toy turbulence model for point sources. We note the primary distortion given our chosen parameters is geometric distortion.}}
\label{fig: Fig12}
\end{figure}

\rev{\subsection{Decompositions for Pixel-wise PSFs}
\label{sec: ex3}
While the two previous examples illustrate the cases of binary masks, we now use a toy model of imaging through turbulence to demonstrate a case of a PSF per-pixel, requiring a more general approach than partitioning. Our toy model is Gaussian kernels with random shifts:
\begin{equation}
    h_{\vu}(\vx) \propto \exp\left\{ -\frac{(\vx - \vt(\vu))^T(\vx - \vt(\vu))}{b(\vu)^2} \right\}.
\end{equation}
Here $\vt(\vu)$ is a random vector field comprised of two independent fields $X$ and $Y$ as $\vt(\vu) = [X(\vu), Y(\vu)]^T$.
Furthermore $b$ is a scalar random field which parameterizes the blur. While the exact details of the fields are not critical, we show the result of a grid of point sources distorted by our toy turbulence model in \fref{fig: Fig12}. For this example we primarily stick to mild blur and stronger geometric distortions, demonstrating some utility for motion-blur modeling.
}

\rev{Relying on per-pixel partitioning would require an entire convolution per-pixel and is therefore very computationally expensive. We instead represent each kernel as a sum of bases obtained via PCA. If an accurate PCA representation requires $K \ll M$ basis functions, where $M$ is the number of pixels in an image, this corresponds to a meaningful reduction in complexity. To achieve this, given that we have access to all the PSFs, we can take a PCA over the entire dataset:
\begin{equation}
    \text{PCA}\left( \{ h_{\vu} \}_{\vu \in \R^2} \right) \rightarrow \{\varphi_m, a_{\vu,m} \}_{m = 1, 2, \ldots M}.
\end{equation}
By this, we can encode the information of each PSF into the invariant bases and spatially varying coefficients.}

\begin{figure}[h]
\centering
\begin{tabular}{ccc}
\rev{(a) Oracle} & \rev{(b) Gathering} & \rev{(c) Scattering} \\
\hspace{-1ex}\includegraphics[width=0.3\linewidth]{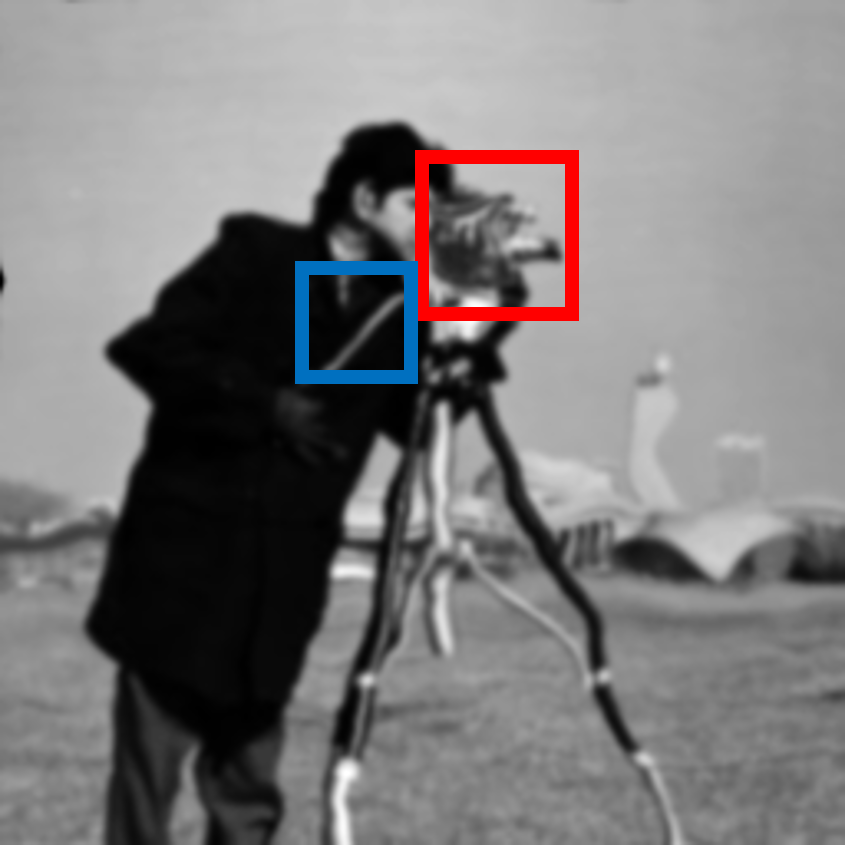}&
\hspace{-1ex}\includegraphics[width=0.3\linewidth]{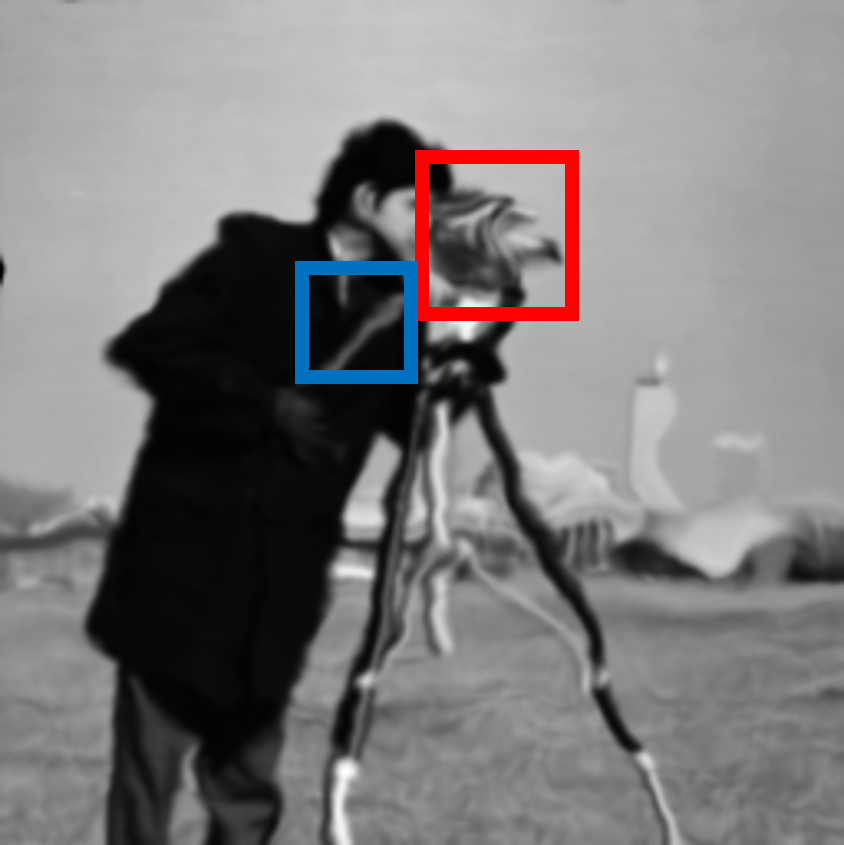}&
\hspace{-1ex}\includegraphics[width=0.3\linewidth]{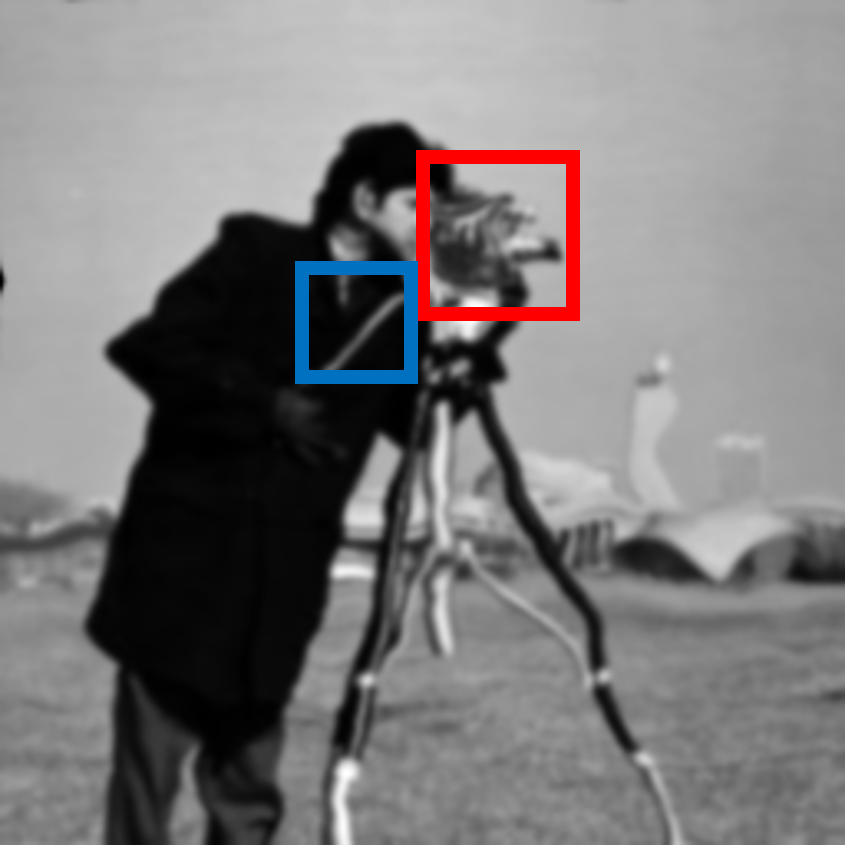}
\end{tabular} \\
\begin{tabular}{cc}
\hspace{-1ex}\includegraphics[width=0.47\linewidth]{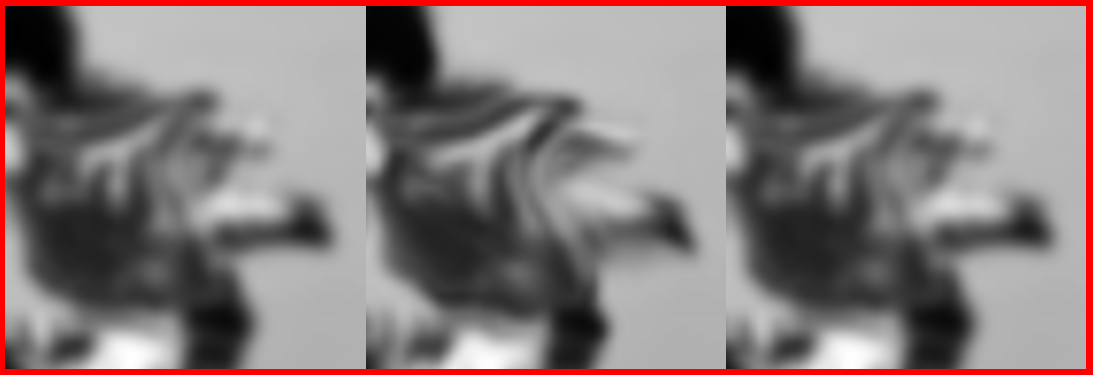}&
\hspace{-1ex}\includegraphics[width=0.47\linewidth]{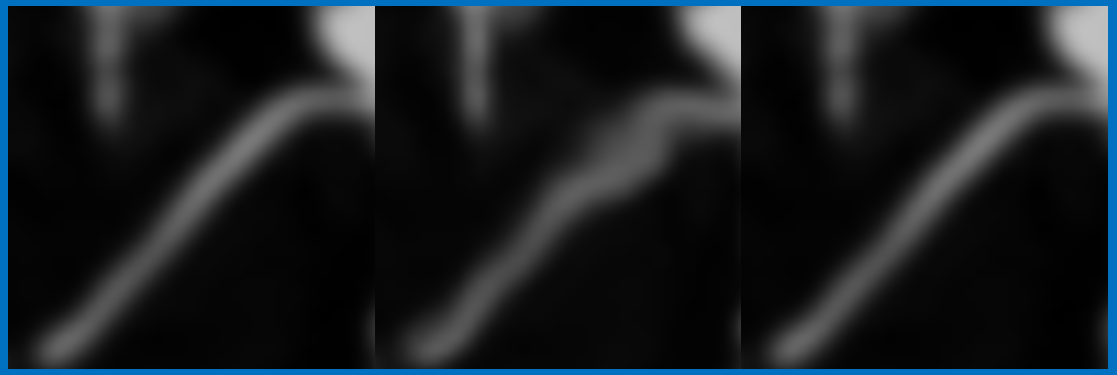}
\end{tabular}
\caption{\rev{A visual comparison of the oracle case and scattering and gathering for our toy turbulence model. We can see the oracle case matches scattering, but has some visual difference with gathering.}}
\label{fig: Fig13 result}
\end{figure}

\rev{Continuing with our toy turbulence example, we can (tediously) simulate the ``oracle'' by using a partition per-pixel. We regard this as the correct result. We then apply the scattering and gathering approximations, facilitated by PCA, and see how they compare to the oracle. The comparison is shown in \fref{fig: Fig13 result} where we can see that scattering (product-convolution) matches the oracle while gathering (convolution-product) has some errors, particularly near boundaries.}

\section{Discussions}
As we write this paper, two underpinning questions are constantly asked: (1) what utility does it bring? (2) How does it affect how we solve an inverse problem? In this section, we briefly share our findings.

\rev{\subsection{Which Decomposition Should I Choose?}
In the examples presented, we have considered both partition and modal interpretations. We consider the following as guidelines regarding which to use:
\begin{enumerate}
    \item \textbf{Partition-like decomposition:} Suitable for a small amount of invariant kernels across the image.
    \item \textbf{Modal decomposition:} Appropriate for systems which require a PSF per-pixel.
\end{enumerate}
}

\rev{The partitioning approach can represent the scene with no error. To illustrate this, consider the superposition integral
\eqref{eq: I varying}. Since we can partition this any way we like, we can partition it in a way such that each partition corresponds to an invariant convolution. For partitions $\{P_m\}_{m=1,2,\ldots,M}$ we can write
\begin{equation}
\int_{\rev{\R^2}} h(\vx, \vu) J(\vu) d\vu = \sum_{m=1}^M \int_{\rev{P_m}} h_m(\vx - \vu) J(\vu) d\vu
\end{equation}
where $h_m(\vx - \vu)$ is the kernel associated with $P_m$. We utilized partition-like decompositions in Sections \ref{sec: ex1} and \ref{sec: ex2}.}

\rev{Although we can consider a per-pixel partition (as done for the oracle of Section \ref{sec: ex3}), it is often computationally infeasible. In these cases it is often preferable to represent each kernel as a sum of bases, such as a modal representation (e.g. PCA). Having previously discussed the computational benefit of a PCA decomposition, we introduce another interesting question, where can error arise in partition-like or modal decompositions? We highlight two errors which can arise in either case:
\begin{enumerate}
    \item \textbf{Using a sub-optimal basis:} If one uses a poor basis, there may be errors representing each $h(\vx,\vu)$ faithfully. 
    \item \textbf{Using the wrong model:} If one uses the wrong model the result will be wrong. The degree of mismatch is related to the manner in which the kernel changes.
\end{enumerate}
}

\rev{First, we illustrate how a sub-optimal basis can impact our representation. For non-negative functions $J(\vu)$ and $h(\vx, \vu)$, we can compute the relative approximation error for a scattering representation as follows:
\begin{align}
    \calL &= \left\vert\int_{\R^2} J(\vu) \left( h(\vx, \vu) - \sum_m a_{\vu, m}\varphi_m(\vu - \vx) \right)\right\vert d\vu \notag \\
    &\leq \int_{\R^2} J(\vu) \left\vert h(\vx, \vu) - \sum_m a_{\vu, m}\varphi_m(\vu - \vx)\right\vert d\vu, \label{eq: app error}
\end{align}
with a similar result for a gathering approximation.
Inspecting the difference term in \eqref{eq: app error}, we note this is just the error in basis representation. Thus, the upper bound of the error is related to the basis representation error.
}

\rev{For a partition-like decomposition, this error arises from choosing the wrong function $\varphi_m$ for partition $P_m$. This may stem from approximating a slowly varying kernel as an invariant one over a partition. For a modal decomposition, each kernel must be captured by the basis decomposition. We note that if all kernels come from the same distribution, a bound can likely be placed on the decomposition and one can say with confidence that the approximation holds.}

\begin{figure}[h]
\centering
\begin{tabular}{cc}
\hspace{-1ex}\includegraphics[width=0.45\linewidth]{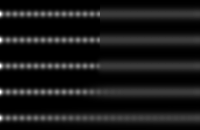}&
\hspace{-2ex}\includegraphics[width=0.45\linewidth]{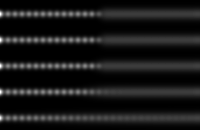}\\
\small{(a) Gathering} & \small{(b) Scattering}
\end{tabular}
\caption{\rev{Visualizing when gathering can approximate scattering. Each row increasingly considers a more slowly changing PSF. Although we do not show it here, the error between the two outputs shown monotonically decreases.}}
\label{fig: Fig14}
\end{figure}

\rev{The other noted listed possible source of error is incorrectly choosing scattering or gathering. Although we have already discussed their mutual exclusivity, consider their equality:
\begin{align}
\sum_{m=1}^M a_{\vx,m} (\varphi_m \rev{\ast} J)(\vx) \overset{?}{=} \sum_{m=1}^M \bigg( \varphi_m \rev{\ast} (a_m \odot J) \bigg)(\vx) \label{eq: gath_scat_eq}.
\end{align}
Are there any conditions which scattering and gathering are \emph{approximately} equal? To illustrate a possibility, we provide a version of the example from Section \ref{sec: ex1} in \fref{fig: Fig14}, where the kernel becomes increasingly invariant as we move toward the bottom of the five rows. The top row corresponds to a sharp transition (equivalent to a partition), while each increasing row uses a more gradual transition. 
}

\rev{As the coefficients become nearly invariant over the support of the kernels, the error in choosing the wrong model lessens. Mathematically, as the coefficients approach approximate invariance (relative to kernel support size) we can justify factoring out the coefficients $a_m$ on the RHS of \eqref{eq: gath_scat_eq}. Thus, the two become increasingly close, with them being exact when we achieve effectively invariant convolution.
}

\rev{The choice in a partition or modal decomposition and the source of errors is highly contextual, depending on various characteristics of the problem. Furthermore, there is nothing restricting us from blending approximations (partition and modal) and scattering and gathering depending on the problem. One such example is adaptively choosing the weights in a blended partition and modal fashion \cite{Alger_2019_a}. Thus, researchers have some flexibility in choosing the proper approximation for their problem.}

\subsection{Utility: Simulating atmospheric turbulence}
\rev{Scattering can be applied to} the simulation of atmospheric turbulence. The latest turbulence simulators, based on phase-over-aperture \cite{Chimitt_2020_OpEng}, phase-to-space (P2S) transform \cite{Mao_2021_ICCV}, and dense-field P2S \rev{(DF-P2S)} \cite{Chimitt_2022_a}, are all using the \emph{gathering} equation, \rev{having some mismatch with nature.}

Table~\ref{tab:text} shows a comparison between the \rev{DF-P2S} simulator \cite{Chimitt_2022_a}, and a new simulator implemented using the scattering equation.\footnote{The new simulator contains a few other modifications including expanding the parameter space, and expanding the kernel supports. However, the biggest change is the adoption of the scattering equation.} Our testing dataset is based on the text recognition dataset released \rev{in} the UG2+ challenge \cite{UG2}. We tested three image reconstruction models: TSRWGAN \cite{Jin_2021_NatureMI}, ESTRNN \cite{Zhong_2022_a}, and TMT \cite{Zhang_2022_arXiv}. We report the recognition accuracy in terms of CRNN / DAN / ASTER from the restored images.

\begin{table}[h]
\centering
\small
\caption{Comparison on the turbulence-text dataset.}
\setlength{\aboverulesep}{0pt}
\setlength{\belowrulesep}{1pt}
\resizebox{0.46\textwidth}{!}{
\begin{tabular}{ccc}
\toprule
 Model &  Simulator in \cite{Chimitt_2022_a}  &  New simulator \\
  & \rev{CRNN / DAN / ASTER} & \rev{CRNN / DAN / ASTER} \\
\midrule
TSRWGAN & 57.75 / 71.45 / 73.10  & 60.30 / 73.90 / 74.40 \\
ESTRNN & 84.50 / 96.25 / 95.45   & 87.10 / 97.80 / 96.95 \\
TMT & 79.25 / 85.20 / 88.00  & 80.90 / 87.25 / 88.55 \\
\bottomrule
\end{tabular}}
\label{tab:text}
\end{table}

As we can see in this table, the new simulator indeed offers a sizable amount of improvement over the previous simulator. Since the change from the gathering to the scattering equation attributes to a substantial portion of the simulator update, the utility of our study is evident.

\subsection{We have an inverse problem, which model to use?}
Solving an inverse often requires an optimization. For a spatially varying blur problem, the typical formulation is
\begin{equation}
\widehat{\mJ} = \argmin{\mJ}  \;\; \|\mI - \mH\mJ\|^2 + \lambda R(\mJ),
\label{eq: inverse main}
\end{equation}
for some regularization functions $R(\mJ)$. \rev{Since we know that the scattering formulation applies to image formation, we can instead write this as}
\begin{equation}
\widehat{\mJ} = \argmin{\mJ}  \;\; \left\|\mI - \left(\sum_{m=1}^M \mH_m\mD_m^{\rev{(s)}}\right)\mJ\right\|^2 + \lambda R(\mJ).
\label{eq: inverse scatter}
\end{equation}
This problem is extremely difficult to solve. \rev{Even if $\mD_{m}^{\vu}$ is binary, we cannot solve for individual regions and combine them. Furthermore, the blurs $\mH_m$ are summed which further complicates the problem. Although we accurately describe the forward model, the inverse problem becomes challenging.}

On the other hand, the gathering equation will give us
\begin{equation}
\widehat{\mJ} = \argmin{\mJ}  \;\; \left\|\mI - \left(\sum_{m=1}^M \mD_m^{\rev{(g)}}\mH_m\right)\mJ\right\|^2 + \lambda R(\mJ).
\label{eq: inverse gather}
\end{equation}
If $\mD_m^{\rev{(g)}}$ is binary (as in Nagy and O'Leary \cite{Nagy_1998_a}), we can partition the image into $M$ smaller regions and solve them individually or through overlapping partitions such as in \cite{Fornasier_2010_a} \rev{or by recurrent neural networks with spatially varying weights as in \cite{Ren_2022_a}}. The parallelism offered by the model is computationally appealing, and it has been proven useful. \rev{The caveat is that there is a model mismatch, and thus is a proxy to the original problem. However, if one can justify an assumption of approximate spatial invariance, one can utilize the fact that gathering may approximate scattering.} In a general sense for arbitrary $\mD_m^{\rev{(g)}}$, any solution we obtain, regardless if we can prove convergence of the optimization algorithm, will not resemble the true solution to \eref{eq: inverse main}.

Our advice to practitioners, when solving an inverse problem related to the spatially varying blur, is to have the sense of awareness for this kind of mismatch. For deep neural networks, the mismatch is often less of an issue when the capacity of the network is large enough. However, the training data needs to capture enough of the physics in order to generalize well.

\section{Conclusion}
\rev{The approximations of} gathering and scattering for a spatially varying blur \rev{were discussed in this work.} They are mutually exclusive, in the sense that if one is the exact representation of the original blur, the other one can only be an approximation. They become identical if the underlying blur kernel is spatially invariant. \rev{In} summary, we recognize the following key points:

\textbf{Gathering has an origin of image filtering.} It is an effective approach to speed up the spatially varying blur via a small set of \rev{invariant} blurs. The approach is to filter the image first, and then combine the filtered images through a pixelwise mask. Gathering offers better edge awareness in tasks such as image denoising.

\textbf{Scattering is originated from light propagation physics.} It is an accurate description according to the scalar diffraction theory in Fourier optics. The approach is to weigh the images and then perform filtering afterward. Scattering is \emph{the} model for describing how light propagates through a random medium.

\begin{IEEEbiography}[{\includegraphics[width=1in,height=1.25in,clip,keepaspectratio]{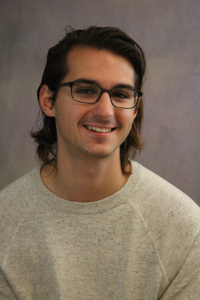}}]
{Nicholas Chimitt}
(S'17--M'23) received the B.Eng. degree in Electrical and Computer Engineering from the Purdue University Northwest, Westville, IN in 2017 and the Ph.D. degree in Electrical Engineering from the Purdue University, West Lafayette, IN in 2023 where he is currently a postdoctoral researcher.

His research works have been published at top computer vision conferences such as CVPR/ICCV, and journals such as the IEEE Transactions on Computational Imaging and SPIE Optical Engineering. He is a co-author of \emph{Computational Imaging through Atmospheric Turbulence}, Now Publisher 2023. His research interests include computational imaging and imaging through atmospheric turbulence.
\end{IEEEbiography}

\begin{IEEEbiography}[{\includegraphics[width=1in,height=1.25in,clip,keepaspectratio]{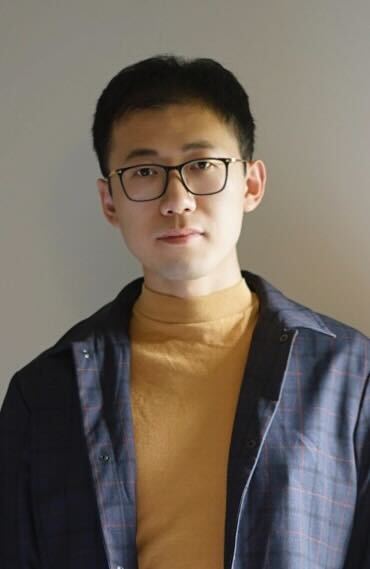}}]
{Xingguang Zhang}(S'20) received a B.E. degree in Optoelectronic Information Science and Engineering from Zhejiang University in 2018, and an M.S. in Electrical and Computer Engineering from Purdue University in 2021. He joined the i2Lab in 2021 as a Ph.D. student, and his research area includes turbulence modeling and mitigation, generative models, domain adaptation, high dynamic range imaging, and machine learning for video restoration. 

His research works have been published at top computer vision conferences, such as CVPR/ICCV, and journals, such as the IEEE Transactions on Computational Imaging. He co-organized the UG2+ workshop at the CVPR 2023 and 2024, he also serves as a reviewer for international conferences and journals, including ICPR, ICME, ECCV, and IEEE TPAMI.
\end{IEEEbiography}

\begin{IEEEbiography}[{\includegraphics[width=1in,height=1.25in,clip,keepaspectratio]{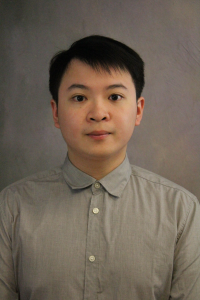}}]
{Yiheng Chi} received the B.S. degree in 2018 in computer engineering from Purdue University, West Lafayette, IN, USA, where he is currently working toward the Ph.D. degree. His research interests include photon limited imaging, imaging on mobile devices, image and video restoration, denoising, computer vision, and machine learning. He is also a regular reviewer for international conferences and journals, including CVPR, ICIP, IEEE OJ-SP, IEEE TIP, IEEE JSTSP, IEEE TCAS-I, Signal Processing: Image Communication, and Optics Express.
\end{IEEEbiography}

\begin{IEEEbiography}[{\includegraphics[width=1in,height=1.25in,clip,keepaspectratio]{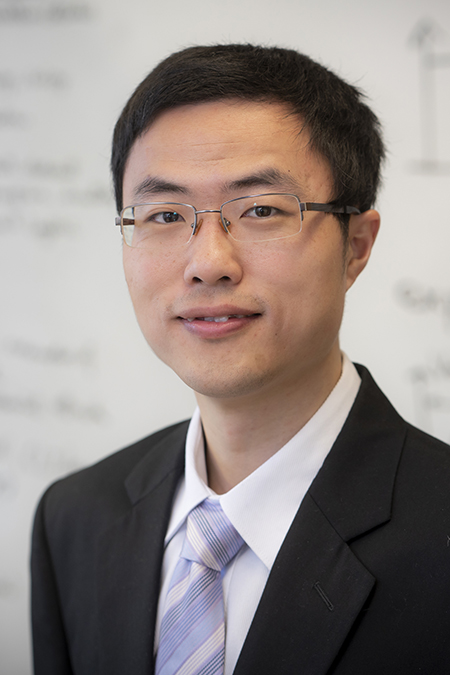}}]
{Stanley H. Chan}
(S'06--M'12--SM'17) received the B.Eng. degree in Electrical Engineering from the University of Hong Kong in 2007, the M.A. degree in Mathematics, and the Ph.D. degree in Electrical Engineering from the University of California at San Diego in 2009 and 2011, respectively. From 2012 to 2014, he was a postdoctoral research fellow at Harvard University. He joined Purdue University, West Lafayette, IN in 2014, where he is currently an Elmore Associate Professor of Electrical and Computer Engineering.
Dr. Chan is a recipient of the IEEE Signal Processing Society Best Paper Award 2022 and the IEEE International Conference on Image Processing Best Paper Award 2016. At Purdue, he is a recipient of multiple teaching awards. He is the author of a popular undergraduate textbook \emph{Introduction to Probability for Data Science}, Michigan Publishing 2021. He is a co-author of \emph{Computational Imaging through Atmospheric Turbulence}, Now Publisher 2023. His research interests include single-photon imaging, imaging through atmospheric turbulence, and computational photography.

He is currently serving as a Senior Area Editor of the IEEE Transactions on Computational Imaging (AE 2018-2022, SAE 2022-now), and a Senior Area Editor of the IEEE Open Journal on Signal Processing (2022-now). He is an advising member of the IEEE Signal Processing Society Technical Committee in Computational Imaging. He was an elected member of the CI-TC in 2015-2020 and an elected member of the IVMSP-TC in 2023-2025.
\end{IEEEbiography}

\bibliographystyle{IEEEbib}
\bibliography{refs}
\end{document}